\documentclass{aa}
\usepackage{epsfig}
\usepackage{amsmath}
\usepackage{txfonts}
\usepackage{natbib}
\usepackage{bm}

\begin{document}

\title{Formation of coronal rain triggered by impulsive heating associated with magnetic reconnection}
\titlerunning{Coronal rain formation by magnetic reconnection}
\authorrunning {Kohutova et al.}

\author{P. Kohutova\inst{1,2}, E. Verwichte\inst{3} \and C. Froment\inst{1,2}}
\institute{Rosseland Centre for Solar Physics, University of Oslo, P.O. Box 1029, Blindern, NO-0315 Oslo, Norway\\
\email{petra.kohutova@astro.uio.no}
\and
Institute of Theoretical Astrophysics, University of Oslo, P.O. Box 1029, Blindern, NO-0315 Oslo, Norway
\and
Centre for Fusion, Space and Astrophysics, Department of Physics, University of Warwick, Coventry CV4 7AL, UK\\}
\date{Received; accepted}

\abstract
{Coronal rain consists of cool plasma condensations formed in coronal loops as a result of thermal instability. The standard models of coronal rain formation assume that the heating is quasi-steady and localised at the coronal loop footpoints.} 
{We present an observation of magnetic reconnection in the corona and the associated impulsive heating triggering formation of coronal rain condensations.}
{We analyse combined SDO/AIA and IRIS observations of a coronal rain event following a reconnection between threads of a low-lying prominence flux rope and surrounding coronal field lines.}
{The reconnection of the twisted flux rope and open field lines leads to a release of magnetic twist. Evolution of the emission of one of the coronal loops involved in the reconnection process in different AIA bandpasses suggests that the loop becomes thermally unstable and is subject to the formation of coronal rain condensations following the reconnection and that the associated heating is localised in the upper part of the loop leg.}
{In addition to the standard models of thermally unstable coronal loops with heating localised exclusively in the footpoints, thermal instability and subsequent formation of condensations can be triggered by the impulsive heating associated with magnetic reconnection occurring anywhere along a magnetic field line.}

\keywords{Magnetohydrodynamics (MHD) -- Sun: corona -- Sun: magnetic fields -- magnetic reconnection}

\maketitle

\section{Introduction}

 \begin{figure*}
\includegraphics[width=43pc]{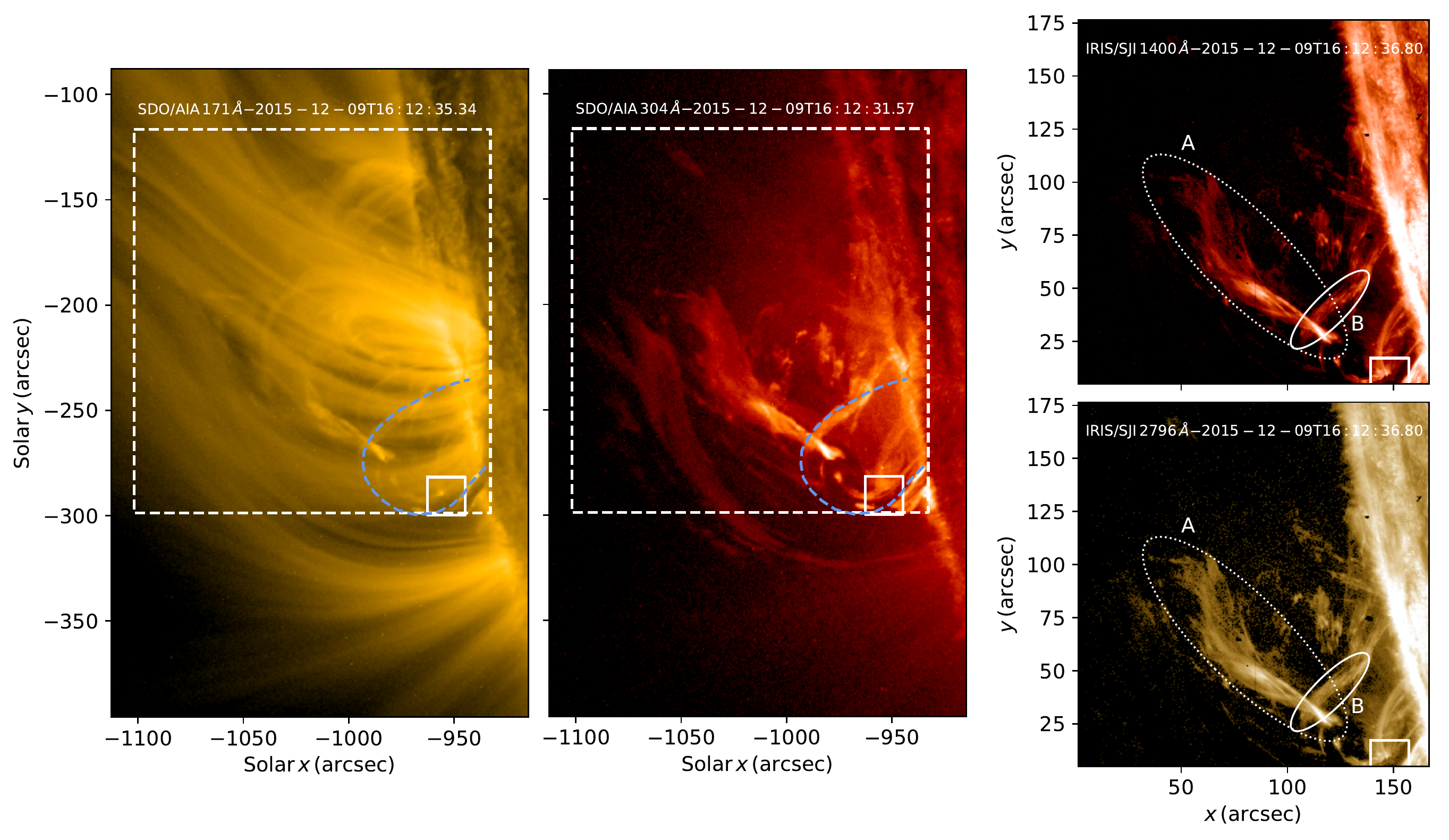}
\caption{Snapshots of the studied active region observed at 16:12:35 in AIA 171 {\AA} (left), AIA 304 {\AA} (middle) and IRIS SJI 1400 {\AA} (top right) and 2796 {\AA} (bottom right), with the dashed line in the AIA images outlining the IRIS FOV. The snapshots show untwisting surge material (A), coronal rain condensations forming in the thermally unstable loop (B), and the active region prominence observed at the limb. The blue dashed line outlines the axis of the coronal loop bundle that turns thermally unstable after the reconnection. The solid square indicates the reconnection region near one of the prominence legs. Animation of this figure is available.}
\label{fig:context}
\end{figure*}

Coronal rain consists of numerous small plasma condensations falling towards the solar surface guided by the magnetic field lines. Coronal rain is a direct consequence of thermal instability \citep{field_thermal_1965} likely to occur in loops that have heating concentrated near their footpoints \citep{karpen_are_2001, muller_dynamics_2003}. The footpoint-localised heating leads to catastrophic cooling of the plasma at coronal heights and the subsequent formation of cool and dense condensations with temperatures ranging from transition region to chromospheric.  It is commonly observed in active region coronal loops \citep{antolin_observing_2012}, suggesting that it plays an important role in the chromosphere-corona mass cycle. Furthermore, condensation plasma remains strongly coupled to the local ion population which means that coronal rain is an excellent tracer of coronal magnetic field and of the fine-scale structure of coronal loops. Recent observations of transverse oscillations in coronal rain \citep{antolin_transverse_2011, kohutova_analysis_2016, verwichte_excitation_2017} highlight its seismological potential. It has become apparent that there are two classes of coronal rain linked to its formation mechanism: quiescent coronal rain and coronal rain formed in the aftermath of solar flares. Even though the fundamental physical processes responsible for the formation of both types of coronal rain are the same, their detailed characteristics are different. Quiescent coronal rain is typically observed in long-lived active region coronal loops. It is caused by the loop footpoint heating and occurs with more gradual mass loading  \citep[e.g.][]{antolin_observing_2012, antolin_multithermal_2015}. This type of coronal rain is more common, as most active region coronal loops are out of hydrostatic equilibrium alternating between heating and cooling phases \citep{aschwanden_modeling_2001} and therefore likely to be subject to thermal instability at some point. Coronal rain associated with solar flares is formed in post-flare coronal loops and is believed to be triggered by the concentrated heating deposited by the non-thermal electron beams. Post-flare coronal rain events are more violent and the rate of mass and energy exchange between the chromosphere and the corona is much greater \citep{scullion_observing_2016}. However, due to the unpredictable nature of solar flares, the multi-instrument observations of flare-driven rain are relatively infrequent, which means its detailed properties are not as well studied as those of its quiescent counterpart. Recent numerical simulations, however, suggest that electron beam heating on its own might not be sufficient to trigger coronal rain formation. It is therefore likely that our model of flare-driven coronal rain is incomplete \citep{reep_electron_2018}.

The magnetic reconnection results in a rearrangement of the magnetic field topology and is responsible for some of the most violent events in the solar corona including solar flares, prominence eruptions, and coronal mass ejections \citep{priest_magnetic_2000, aschwanden_physics_2005}. One of its defining features is the rapid conversion of magnetic energy into kinetic energy and heat \citep{parker_solar_1963}. Due to the lack of methods that can be used to directly determine the topology of the coronal magnetic field on short timescales without relying on the extrapolation of the photospheric magnetic fields (as the magnetic fields immediately prior to eruption are expected to be highly non-potential), the evolution of the magnetic field topology during reconnection events is usually constrained by analysing the changes in morphology of the coronal structures, for example in coronal loops seen in imaging data at extreme ultraviolet (EUV) wavelengths. These indirect signatures include bidirectional reconnection outflows \citep{mckenzie_xray_1999, wang_direct_2007, tian_imaging_2014}, plasmoid ejection \citep{shibata_hot_1995, liu_plasmoid_2013}, the presence of current sheets \citep{mckenzie_xray_1999, sui_evidence_2003}, hard X-ray emission \citep{masuda_looptop_1994}, and the generation of fast-mode MHD waves, both propagating \citep{li_quasi_2018} and standing in the surrounding coronal structures \citep{white_antiphase_2013}. Main observational challenges associated with reconnection events are due to the small spatial and temporal scales on which the key processes occur. The detailed understanding of reconnection events in the corona therefore relies on combining evolution observed in coronal wavelengths with a number of different indirect signatures, only a few of which might be present at the same time. 

 \begin{figure*}
\includegraphics[width=40pc]{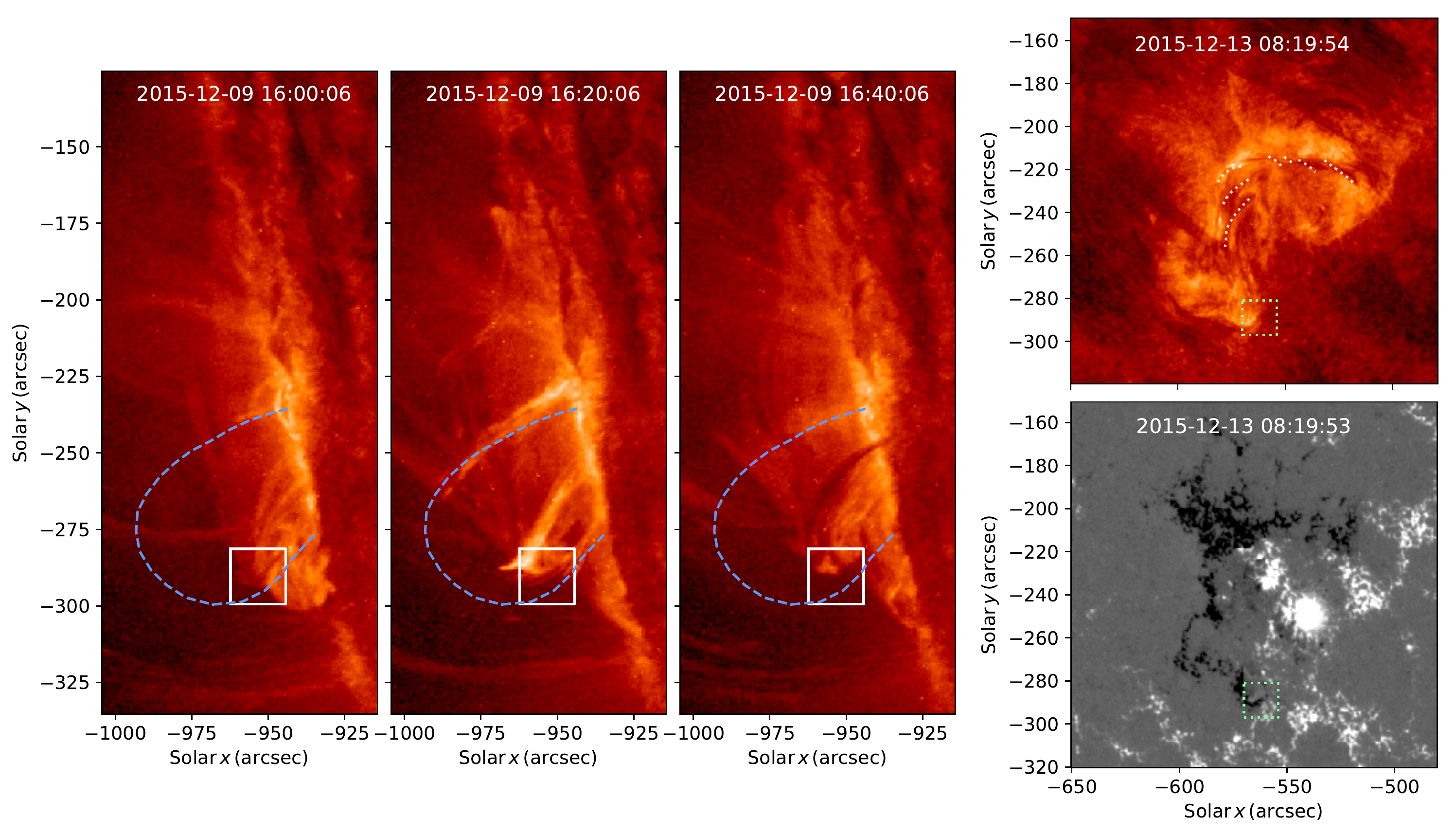}
\caption{Left to right: AIA 304 {\AA} snapshots showing the prominence at the limb immediately before the reconnection occurring at 16:02 UT, shortly after, and in a new quasi-steady state. The solid square indicates the location of the reconnection site. The blue dashed line outlines the axes of the coronal loop bundle that becomes thermally unstable. After a few days, the prominence is still visible on-disc as a filament (top right). The white dotted lines outline the filament turns. The HMI magnetogram (bottom right) shows that the flux rope lies along a polarity inversion line with one leg embedded in a flux cancellation region (green dotted square).}
\label{fig:prominence}
\end{figure*}

One of possible reconnection events occurring in the solar corona is triggered by the interaction between an erupting filament (or prominence, if observed off-limb in emission) and the surrounding coronal structures. The erupting filament typically reconnects with a closed magnetic structure such as a nearby coronal loop \citep{driel_coronal_2014} or with an open magnetic field line via interchange reconnection \citep{baker_signatures_2009, yang_interchange_2015}. As the magnetic structure in filaments is usually subject to a non-zero magnetic twist as the cool and dense prominence plasma is confined in a magnetic flux rope, their eruptions are likely to result in the magnetic twist release \citep{xue_observing_2016}. Filament eruptions are often associated with the onset of a hot jet or a surge of cool plasma ejected via the slingshot effect of the snapping magnetic field lines where the heating is a result of a reconnection-induced Joule dissipation \citep{yokoyama_magnetic_1995, sterling_small_2015, li_blowout_2017}.

Coronal rain formation has been previously assumed to be caused exclusively by localised heating concentrated at the loop footpoints. The link between magnetic reconnection and formation of plasma condensations in the corona has been investigated in terms of the reconnection induced topology changes \citep{kaneko_reconnection_2017, li_coronal_2018}. \citet{li_coronal_2018} studied the onset of condensation downflows reminiscent of a coronal rain event triggered by the magnetic topology change due to magnetic reconnection. In that event a condensation region was formed in a dip in an open magnetic field line most likely via the standard footpoint-heating mechanism. Once the field line reconnected with an underlying closed loop structure, the condensation material was no longer trapped and falls towards the solar surface in a rain-like manner. In that case it is therefore the downfalling motion of the condensations that is triggered by the magnetic reconnection which breaks the dip supporting the condensations, but not the condensation formation itself. Given that the values of plasma beta typically estimated for large condensation regions are of the order of one, the mass of the condensation plasma trapped in a magnetic dip can also deform the magnetic field and cause its downward motion and the subsequent reconnection \citep{hillier_numerical_2012}. 

The standard model of coronal rain formation further assumes the footpoint concentrated heating to be sustained and quasi-steady (i.e. acting on timescales much shorter than the radiative cooling timescale). This is found to lead to the thermal non-equilibrium scenario, where a coronal loop undergoes repeated heating-condensation cycles, as evidenced by both observations \citep{froment_evidence_2015, auchere_coronal_2018} and modelling \citep{fang_coronal_2015}. In this scenario the characteristic coronal rain formation timescale is long (of the order of hours), which is equivalent to the time it takes for the loop to gradually fill with evaporated plasma and for the density in the loop to increase sufficiently to trigger thermal instability. The effect of impulsive one-off events providing isolated sources of heating that lead to transient coronal rain showers has not yet been considered. 

In this work we show that the magnetic reconnection can be directly responsible for the thermal instability onset and subsequent condensation formation through the impulsive localised heating associated with this process. We present an observation of a reconnection of twisted prominence threads with surrounding coronal structures. The associated impulsive heating localised in the upper leg of an adjacent coronal loop leads to coronal rain formation. 

\section{Data}

\begin{figure*}
\includegraphics[width=40pc]{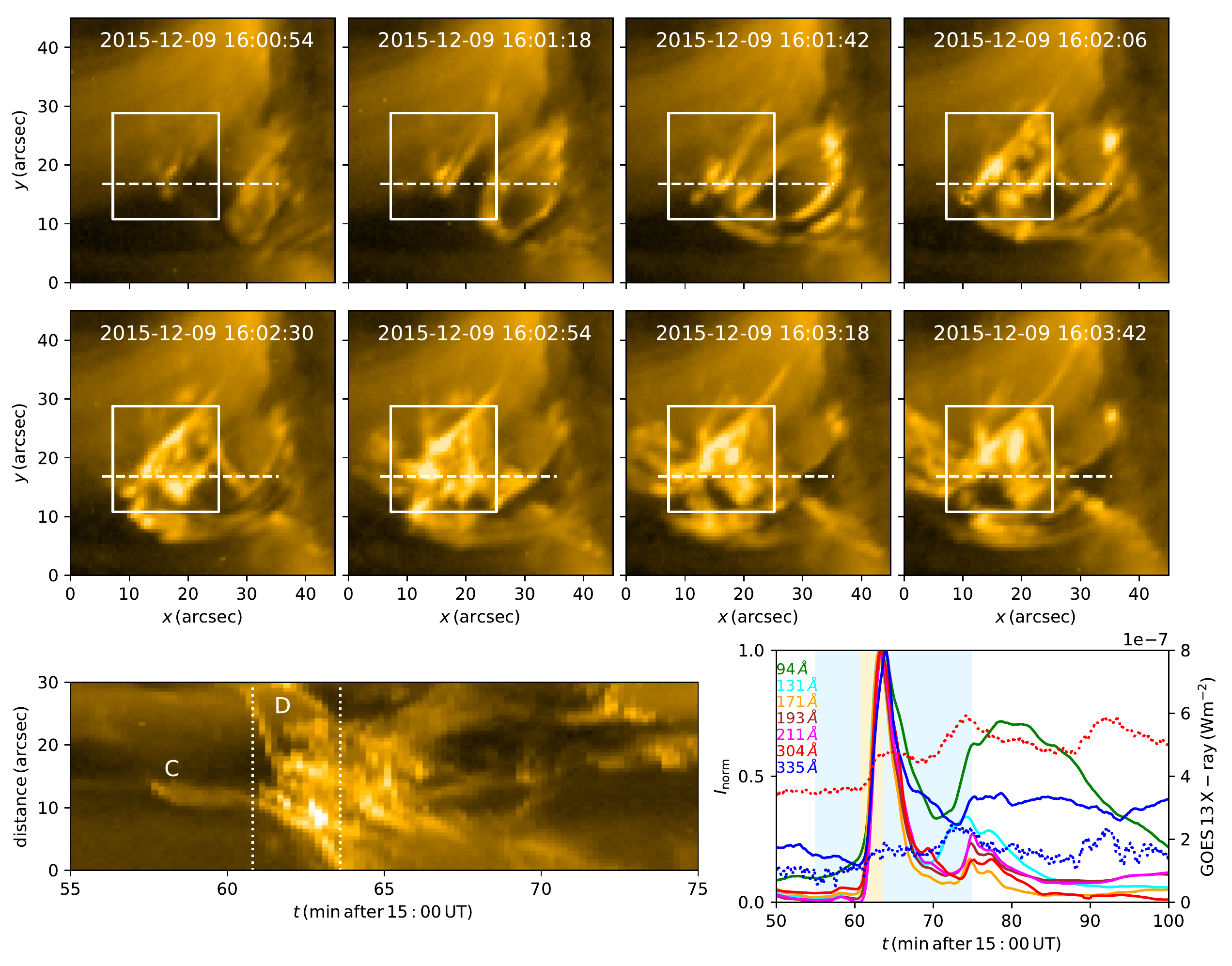}
\caption{Top: Snapshots taken from AIA 171 data covering the time interval 2015-12-09 16:00:54 - 2015-12-09 16:03:42 during which the reconnection occurs zoomed on the reconnection region. The white square marks the reconnection region over which the emission intensities are integrated. Bottom left: Distance–time plot of the approaching low-lying loop-like structures seen in AIA 171. The y-axis corresponds to distance along the horizontal dashed line shown in the snapshots above; the x-axis corresponds to time. The paths of individual structures C and D are indicated in the distance–time plot. Vertical dotted lines give the time range covered by snapshots above. Bottom right: Solid lines show the evolution of normalised emission intensities in individual AIA bandpasses integrated over reconnection region. Red and blue dotted lines correspond to the evolution of solar X-ray irradiance as measured by GOES 13 in the XRS long-wavelength channel (0.1-0.8 nm), and in the XRS short-wavelength channel (0.05 - 0.4 nm), respectively. The short-wavelength irradiance has been multiplied by a factor of 10 to improve clarity. The blue shaded region corresponds to the time range shown in the distance–time plot on the left. The yellow shaded region corresponds to the time range covered by the snapshots above. The duration of the impulsive phase associated with the reconnection corresponding to the increasing emission in all AIA channels is $\sim$ 3 minutes.}
\label{fig:MR}
\end{figure*}

\begin{figure}
\includegraphics[width=20pc]{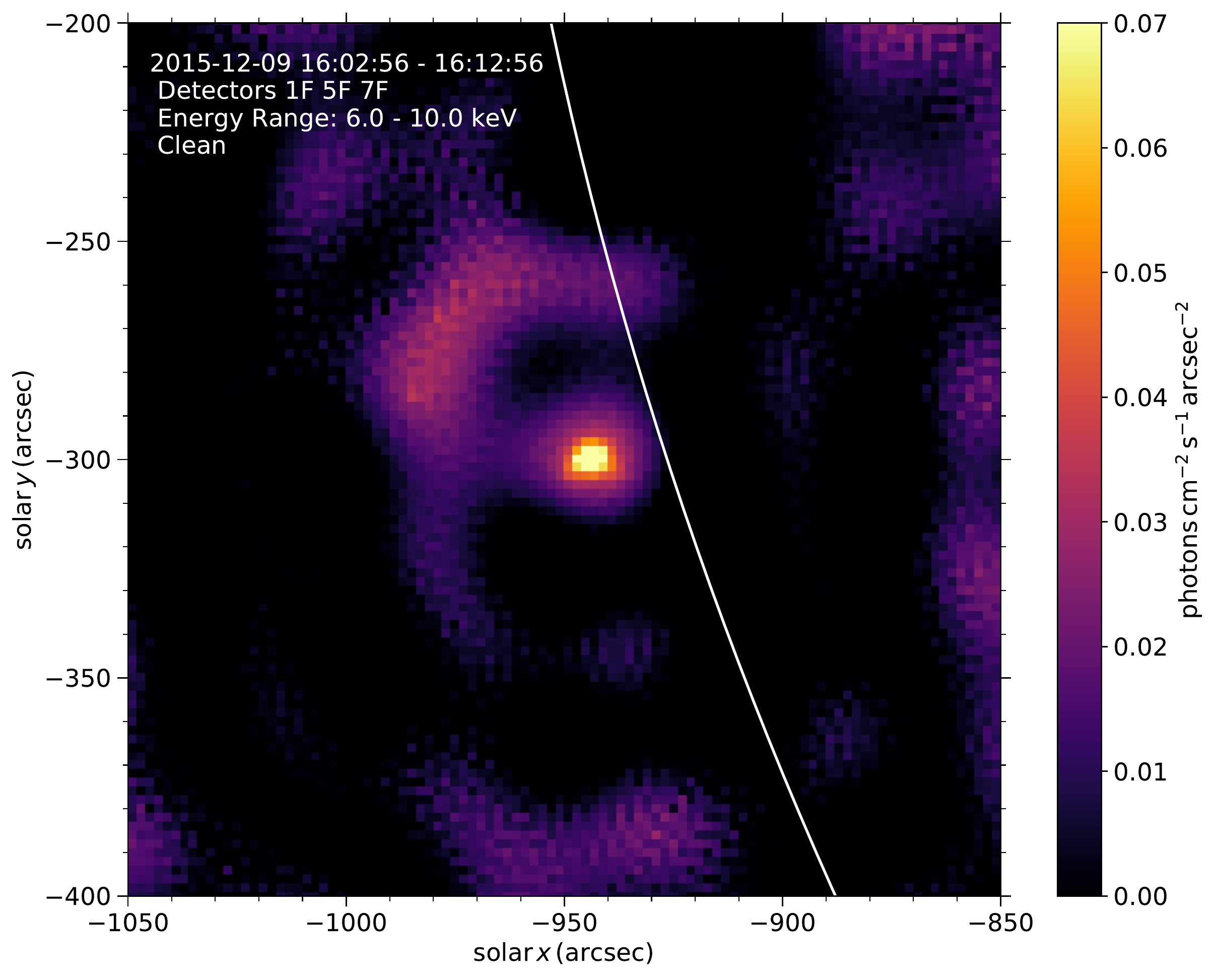}
\caption{Reconstructed RHESSI image corresponding to the X-ray emission during the event in the 6 - 10 keV range carried out using the CLEAN reconstruction algorithm. The solar limb is shown in white.}
\label{fig:rhessi}
\end{figure}

We used observations taken by the \textit{Interface Region Spectrograph} (IRIS) \citep{depontieu_interface_2014} and by the \textit{Atmospheric Imaging Assembly} (AIA) on board the \textit{Solar Dynamics Observatory} (SDO) \citep{lemen_atmospheric_2012}. An event from 9 December 2015 in NOAA AR 12468 was analysed using imaging and spectral IRIS data in chromospheric and transition region wavelengths and complemented by full-disc imaging SDO/AIA data in coronal wavelengths. We used IRIS level 2 slit-jaw imager (SJI) data taken between 16:12 and 17:11 UT. The SJI data are taken in two passbands, the far-UV (FUV) passband is centred on 1400 {\AA} and is dominated by two Si IV lines formed in the transition region at $\log T = 4.8$ and the near-UV (NUV) passband is centred on 2796 {\AA} dominated by the Mg II K line core formed at $\log T = 4$, with an exposure time of 8 s, cadence of 19 s, and image scale of $0.166 \arcsec$ pixel$^{-1}$. The IRIS observations were taken in very large sit-and-stare mode with maximum field of view (FOV) $167\arcsec \times 174\arcsec$ centred at x,y:$\left[-1017 \arcsec, -209 \arcsec \right]$. IRIS data were retrieved from mission web page\footnote[1]{http://iris.lmsal.com/search}. We also used SDO/AIA EUV imaging data in seven broad passbands covering the temperature range from the transition region to coronal values. Level 1.5 SDO/AIA data have an image scale of $0.6 \arcsec$ pixel$^{-1}$, cadence of 12 s, and were normalised by the exposure time. Required subframes of level 1.5 SDO/AIA data were retrieved using the AIA Cutout Service\footnote[2]{http://www.lmsal.com/get\textunderscore aia\textunderscore data/}.

The observing sequence focuses on the NOAA 12468 active region at the eastern limb, which contains a number of large coronal loops with heights of the order of 100 Mm, including a number of footpoints of transequatorial loops that span into northern hemisphere, as well as several shorter ones (Fig. \ref{fig:context}). Several hours prior to the reconnection event, quiescent coronal rain can be observed forming in the long coronal loops. A low-lying prominence with a clear twisted flux-tube structure spanns across the whole active region and is visible in the AIA 304 channels and in the IRIS FUV and NUV SJI images. The magnetic reconnection involving several prominence threads occurs in the foreground component of the prominence, leading to the change in the morphology of the foreground prominence leg. The bulk of the prominence plasma remains quasi-stable, however, and can be observed several days later in AIA 304 {\AA} on-disk as a filament. The IRIS observing sequence starts at 16:12:36 UT i.e. after the main reconnection event has already taken place. The analysis of the reconnection event itself is therefore limited to SDO/AIA data. 

This IRIS dataset was previously analysed by \citet{schad_neutral_2018}, who focused on the analysis of the kinematics and morphology of coronal rain blobs formed in large coronal loops overarching the active region. However, the analysis excluded the upward-moving surge material and the associated downflows. It also excluded the analysis of the low-lying thermally unstable loop this work is focusing on.

\section{Magnetic reconnection and associated impulsive heating}

\begin{figure*}
\includegraphics[width=40pc]{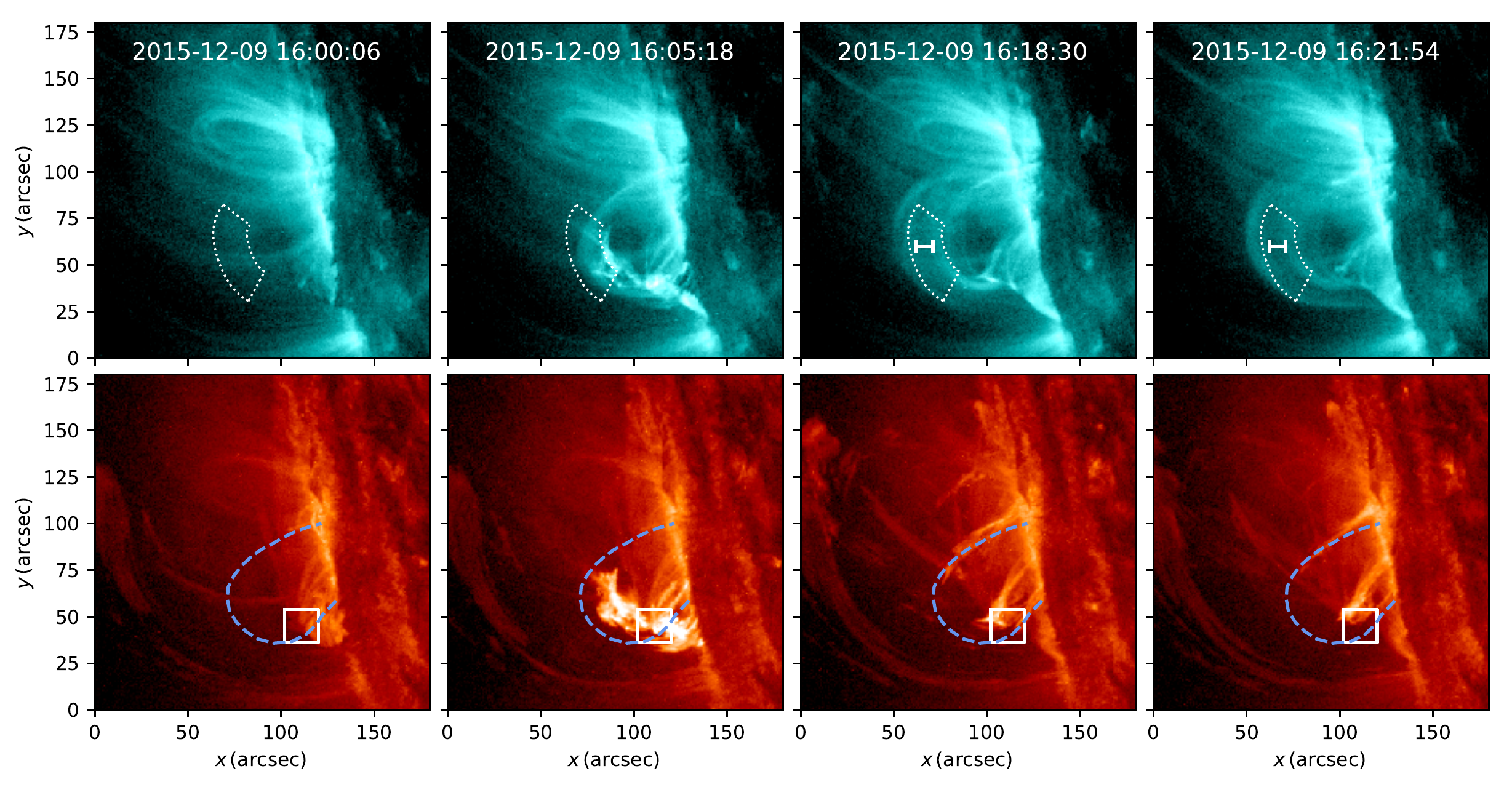}
\caption{Evidence of catastrophic cooling in the studied loop. We plot the snapshots of the loop evolution in the hotter 131 {\AA} channel (top) and cooler 304 {\AA} channel (bottom). The snapshots correspond to the loop not being visible before the reconnection event, to the heating of the loop immediately following the reconnection, the dimming of the central part of the loop in the coronal wavelengths and the coronal rain formation along that field line and to final stage corresponding to the evacuation of the loop following the coronal rain shower (left to right). The dotted line outlines the position of the loop top.The white horizontal bar in the 131 {\AA} snapshot shows the evacuated loop bundle that became thermally unstable. The solid square indicates the location of the reconnection site. The blue dashed line shows the axis of the coronal loop bundle that becomes thermally unstable.}
\label{fig:AIA_loop}
\end{figure*} 

In AIA 304 {\AA} the foreground leg of the prominence appears dynamic and shows a great deal of small-scale motion during the hours preceding the IRIS observing sequence (Fig. \ref{fig:prominence}). This dynamics often leads to the additional braiding of the magnetic field in one prominence leg until it reaches a limit where the prominence magnetic field becomes kink-unstable \citep{hood_kink_1979}. The trigger for reconnection in this region is therefore likely the kink instability resulting from the large amount of twist accumulated in one end of the flux rope, with the helical motion of the ejected material being evidence of the magnetic twist release during the reconnection. As a result of the kink instability the individual flux-rope threads reconnect with the surrounding coronal field lines, both open and closed, as evidenced by the trajectories of material ejected away from the reconnection site, assuming it follows the newly formed magnetic field geometry. The main reconnection event takes place at 16:02 UT. In this scenario the magnetic energy is converted into kinetic energy of the cool prominence plasma ejected away from the reconnection site upwards and into ohmic heating localised in the region shown in Fig. \ref{fig:MR}. The evolution of the emission intensity shows a nearly simultaneous increase in all AIA bandpasses at the reconnection site (Fig. \ref{fig:MR}). The emission immediately follows an approach of two threads within the prominence core, which are labelled C and D in Fig. \ref{fig:MR}. The reconnection region show a series of multiple small-scale brightenings occurring on a timescale of several minutes. The width of the peak in the integrated emission curves is $\sim$ 5 minutes, while the impulsive phase, during which the emission increases in all channels is $\sim$ 3 minutes. This suggests that the main event in fact comprises a formation of multiple current sheets within the reconnection region resulting in a cascade of multiple small-scale reconnection events affecting only a finite fraction of the individual prominence threads. The detailed evolution of the integrated emission in all bandpasses, however, is not studied in detail due to presence of multiple structures along the line of site in the vicinity of the studied loop, as this is likely to result in contamination by the emission of the background plasma.  

The event is also detectable in the soft X-ray RHESSI data (Fig. \ref{fig:rhessi}). The emission is observable in the 6 - 10 keV range above the limb, suggesting that it corresponds to emission from the reconnection site in the corona and not to the footpoint emission due to bremsstrahlung of the accelerated electrons, which lies in the hard X-ray range and is localised in the chromosphere.
 
The surge of cool plasma ejected following the reconnection consists of two components: a freely moving component that shows clear signs of untwisting helical motion that moves along an open field line and eventually disappears from the field of view and a confined component that moves upwards along a closed magnetic loop, subsequently falling back down towards the foot of the newly reconnected flux tube, while exciting a damped transverse oscillation in the loop. The downfalling surge material outlines the newly formed magnetic field configuration consisting of a closed coronal loop with one of the legs threading through the foreground part of the prominence. Following the event, the prominence is still observable with a clearly detectable twist near the prominence footpoints (Fig.  \ref{fig:prominence}) which implies that the eruption was only partial and that most of the prominence large-scale structure, including the magnetic twist, was preserved. Several days later the prominence can still be observed in AIA 304 {\AA} as a filament on-disc. The SDO/AIA 304 {\AA} data clearly shows a significant twist in the prominence, corresponding to at least three complete turns, or equivalently $\Phi = 6 \pi$, thus exceeding the Kruskal–Shafranov criterion for the stability of twisted flux tubes $\Phi_{\mathrm{crit}} = 2 \pi$. This suggests that the prominence was kink-unstable 40 hours after the reconnection event analysed here, after it rotated sufficiently to be clearly observed on-disk. Assuming that the dynamics at the footpoints of the prominence has not changed significantly during these 40 hours, it is reasonable to assume that the kink instability was what triggered the original reconnection event as well. The SDO/HMI magnetogram data taken after the active region has rotated sufficiently to be observed on-disk clearly shows that the prominence lies along the polarity inversion line (Fig.\ref{fig:prominence}). It also suggests that the prominence leg is embedded in a flux cancellation site. Interaction of mixed-polarity surface magnetic fields leading to flux cancellation causes disturbances in the lower layers of the solar atmosphere, which can lead to the additional braiding of the magnetic field in the prominence leg resulting in the kink instability.

\section{Thermal instability and formation of condensations}

The condensation region in the thermally unstable loop appears in the AIA 304 {\AA} bandpass at 16:10 UT. Combining the observations from hotter coronal bandpasses and cool AIA 304 {\AA} provides insight into the thermal evolution of the newly formed coronal loop. As seen in 131 {\AA} emission, the loop first appears in hotter channels immediately following the reconnection event, as it is heated to high temperatures (Fig. \ref{fig:AIA_loop}). In this case the heating is localised in the upper part of the loop leg in the region where the reconnection is taking place. The strong asymmetric heating can be expected to trigger large-scale flows within the loop, leading to the increase in the density of the loop plasma in the upper half of the loop. This results in the acceleration in the radiative loss rate which leads to the loop plasma cooling faster and eventually becoming thermally unstable. At this stage a darkening of the central part of the loop is observable in the  AIA 131 {\AA} bandpass, while the emission in the surroundings continues to increase and no condensation can be observed in AIA 304 {\AA} outside the central part of the bundle. This suggests that only a relatively narrow bundle of field lines actually becomes unstable, with a width of $\sim 8 \arcsec$, or $\sim 5$ Mm. This can be a consequence of the reconnection-induced heating being more spatially concentrated compared with the standard rain formation scenario. Hence the spatially concentrated heating would only affect a fraction of the loop bundle cross section, given that the thermal conduction acts predominantly parallel to the direction of the magnetic field. It could alternatively be a signature of the dependence of the stability of the plasma along single field line on the length of that field line. Previous numerical simulations suggest that the evolution along a given field line and therefore also its thermal stability is strongly dependent on the ratio of the heating scale-height to the field line length \citep{muller_dynamics_2003, fang_coronal_2015}. This can thus lead to a non-uniform evolution across a wide bundle of field lines if their lengths vary. The difference in lengths for a narrow loop bundle, however, might not be large enough to produce observable variation; the first explanation therefore seems more applicable in the studied case. The thermal instability onset in the observed loop eventually results in the formation of the cool cloud of the condensed plasma appearing in the  AIA 304 {\AA} bandpass slightly away from the loop apex, which then falls towards the solar surface along both legs of the loop. This progression from hotter to cooler bandpasses is common observational evidence of the progressive cooling of the plasma seen in both active region loops \citep{ugarte_investigation_2006, ugarte_active_2009} and in bulk of the diffuse coronal plasma \citep{viall_evidence_2012}. It has also been previously observed in coronal loops subject to thermal instability \citep{antolin_multithermal_2015,froment_evidence_2015, kohutova_analysis_2016, auchere_coronal_2018}. The short length of the thermally unstable loop and the larger heating output and therefore greater rate of the evaporation into the loop compared with the quiescent rain scenario result in a shorter instability timescale than typically observed.

%MR-induced rain shows more emission in the hot AIA channels than the quiescent case, especially during the initial phases of the formation process. This is due to the fact that the clumps of the downfalling material observed immediately after the main reconnection event are associated with the reconnection outflows and not formed by gradual cooling of the coronal loop. The condensation material instead moves with much lower speeds as it consists of a cloud of cool plasma formed near the apex of the unstable loop, that slowly starts sliding down along the loop leg due to gravity.  

The off-apex formation of the condensation region, which results in the preferential motion of the coronal rain along the top loop leg, is caused by the asymmetric nature of the heating. In this scenario the impulsive heating is localised in the reconnection site in one loop leg, whereas the classical model of coronal rain formation typically assumes either symmetric footpoint heating or heating with comparable amplitude active at both footpoints. As previously discussed, the impulsive heating caused by reconnection is localised in the middle part of the loop leg, where the peak in the emission in all AIA bandpasses is observed preceding the condensation formation. As the downfalling coronal rain material hits the solar surface, small-scale brightenings can be observed at the impact location (Fig. \ref{fig:AIA_loop}). Similar phenomena linked to coronal rain have been observed in \citet{kleint_detection_2014} and \citet{jing_unprecedented_2016}. They are typically explained as a result of collision of the individual rain blobs with the dense transition region plasma or as a result of compression and the resulting heating of the plasma confined under the rain blobs \citep{muller_dynamics_2003}.  A shock front caused by instability-induced siphon flows transitioning from supersonic to subsonic speeds has also been suggested as a possible explanation \citep{jing_unprecedented_2016}.   

\begin{figure*}
\includegraphics[width=42pc]{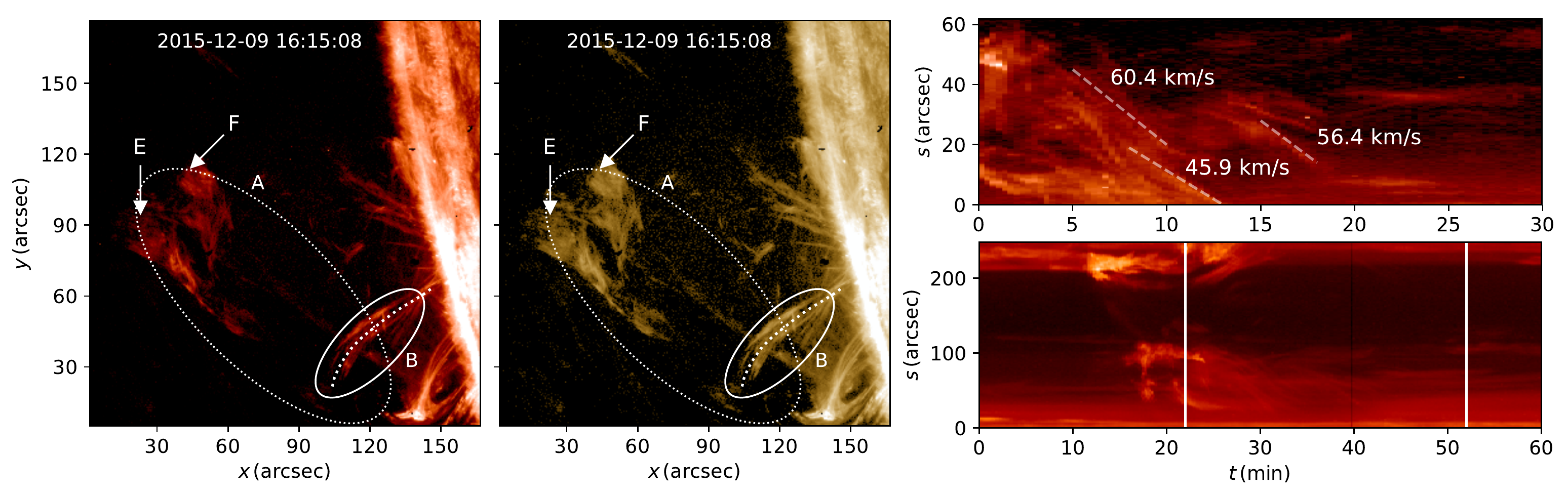}
\caption{IRIS SJI data in Si IV (left) and Mg II K (middle) showing coronal rain condensation formed due to reconnection-induced heating. The surge (A) and the thermally unstable coronal loop with coronal rain (B) are both visible. Also visible are the freely moving (E) and confined surge components (F). The motion of the coronal rain along the loop is shown in the time-distance plot (top right) taken along the dotted line shown in SJI snapshots. The y-axis corresponds to the distance along the loop. Trajectories of three coronal rain condensations with corresponding average speeds are outlined by dashed lines as an example. The evolution along the whole loop in AIA 304 {\AA} shows coronal rain falling down along both loop legs (bottom right). The vertical solid lines in the AIA 304 {\AA} time-distance plot mark the beginning and end of the IRIS observing sequence.}
\label{fig:IRIS_rain}
\end{figure*}

The rain material formed as a result of thermal instability in a low-lying loop should not be confused with the cool plasma ejected during the reconnection. The two different components do not appear to move along the same magnetic field lines, their co-spatiality is just a projection effect and they diverge as they approach the solar surface. The high-speed prominence material is ejected in a single direction only away from the reconnection region, while the condensations formed in the loop are observed falling down along both loop legs (Fig. \ref{fig:IRIS_rain}). The ejected plasma is more multi-thermal and can be observed in cool channels (IRIS FUV and NUV and AIA 304 {\AA}) and in hotter AIA wavelengths, and moves with much higher speeds. This can be explained by the fact that the reconnection and therefore the associated heating is occurring directly in the prominence, so parts of it are heated to coronal temperatures, similarly to the main component of the ejected surge material moving upwards. The condensation material instead moves at much lower speeds as it consists of a cloud of cool plasma formed near the apex of the unstable loop, that slowly starts sliding down along the loop leg, due to gravity (Fig. \ref{fig:IRIS_rain}). We estimate the average projected velocity to be 180 km s$^{-1}$ for the ejected prominence material and 60 km s$^{-1}$ for the coronal rain formed via the condensation process, which is significantly less than the maximum free-fall speed of 150 km s$^{-1}$ estimated for a loop of this size.

\section{Discussion and conclusions}

The reconnection of the twisted prominence threads in the case analysed in this work is triggered by the onset of kink instability. Helical kink instability is a common mechanism responsible for the eruptions of prominences with flux-rope structure \citep{torok_confined_2005, zhang_observation_2012}. In the flux-rope configuration the cool condensation plasma is supported against gravity by helical field lines, while at the same time storing large amounts of magnetic energy. However, the instability and subsequent reconnection is found to affect only a fraction of prominence threads. Even though the release of the magnetic twist is observed in the cool plasma ejected following the reconnection, a major part of the prominence remains observable for several days following the event. The on-disc AIA 304 {\AA} observations also suggest that the magnetic twist supporting the prominence mass remains preserved. We therefore conclude that the event we observed corresponds to a partial prominence eruption. The reconnection is followed by the ejection of cool plasma following two different trajectories, a freely moving component subject to helical motion moving along an open field line and a confined component that moves upwards along a closed magnetic loop while exciting a transverse oscillation in the loop. A detailed analysis of this surge event will be carried out in a forthcoming paper. After several days, the prominence can be observed on-disk as a filament. The SDO/HMI magnetogram data suggests that the prominence is formed along a polarity inversion line and that one of the prominence legs is embedded in a flux cancellation site. Interaction of mixed-polarity surface magnetic fields leading to flux cancellation cause disturbances in the lower layers of the solar atmosphere that can lead to the magnetic field braiding in the footpoints of the magnetic structures. This mechanism is therefore likely responsible for the additional braiding of the magnetic field in the prominence leg leading to the kink instability and subsequent reconnection.

The coronal rain formation, which is a direct result of reconnection-induced localised heating, is more abrupt compared with the commonly observed coronal rain events. The quiescent coronal rain appears in the loop in a more gradual manner and on more prolonged timescales, where the total duration of single coronal rain event often lasts 60 minutes or longer \citep{kohutova_analysis_2016}, with the condensations continuously forming in the loop during this time. The rapid formation in the reconnection-triggered scenario is characterised by a large amount of coronal rain condensations appearing suddenly in the unstable loop (i.e. the condensation formation occurs on shorter timescales than in the quiescent case) and the total duration of the coronal rain shower from the appearance of condensation to the evacuation of the loop is only 20 minutes (Fig. \ref{fig:IRIS_rain}). This is in contrast with the standard model of coronal rain formation which assumes that the heating is sustained for extended periods of time, often leading to quasi-periodic behaviour of the coronal loop consisting of repeating coronal rain events occurring in the same loop on comparable timescales. The coronal rain event studied in this work is transient, however, and does not recur as it is triggered by an impulsive event.

The coronal rain plasma can be distinguished from the prominence material by looking at their trajectories and average speeds. The timescale on which the coronal rain forms following the heating onset is much shorter than for the quiescent scenario; in the studied event condensations appear 10 minutes after the reconnection event, whereas observations of quiescent rain suggests it recurs in the same loop of the order of hours \citep{antolin_observing_2012, kohutova_analysis_2016}. The period of the loop heating-condensation cycle in the quiescent scenario is equivalent to the time it takes for the sustained footpoint heating to refill the loop sufficiently with evaporated plasma to reach the thermally unstable regime, after the loop has been evacuated by the previous coronal rain event. This short timescale for coronal rain formation is likely a consequence of the heating input being much greater than in the quiescent case and of the short length of the studied loop. The 1D numerical simulations suggest that although the loop length is a contributing factor, the heating input is the main factor affecting the coronal rain formation timescale \citep{froment_occurrence_2018}. The thermal instability in the case associated with magnetic reconnection is also more concentrated spatially and only a certain fraction of the loop with a cross section of around 5 Mm width becomes unstable. This implies that the heating that triggers the thermal instability is more localised and only affects a small number of field lines in the loop. As the thermal conduction acts predominantly along the magnetic field, most of the matter and energy transport occurs along the affected field lines. Comparing this to the quiescent scenario, the typical width of the loop bundles observed to undergo condensation formation is around 10-15 Mm \citep{antolin_observing_2012, kohutova_analysis_2016}, and in some cases, reaching 40 Mm \citep{auchere_coronal_2018, froment_multi_2019}.

Off-apex formation of the bulk of the coronal rain material suggest that the location of the condensation region depends on where along the loop is the heating localised, in this case in the upper part of the loop leg around 20 Mm above the footpoint, which is higher than the usual footpoint-concentrated heating. We estimate the radius of the loop to be around 40 Mm (we note that this estimate is subject to uncertainties due to projection effects), which corresponds to the total loop length of 126 Mm, assuming a semi-circular loop. The impulsive heating is therefore localised at ~16 \% of the total loop length. This means that the heating is sufficiently concentrated and the heating scale-height is short compared with the total length of the remaining part of the loop, which is one of the prerequisites for the thermal instability onset \citep[e.g.][]{ muller_dynamics_2003, muller_high_2005}. The sudden increase in temperature at the reconnection site triggers large-scale flows in the loop to compensate for the newly created non-uniform temperature profile along the loop. It has been shown that reconnection occurring in kink-unstable flux tubes leads to the development of large-scale flows directed away from the reconnection site \citep[e.g.][]{haynes_observational_2007}; large-scale redistribution of mass along the analysed loop is therefore very likely following the observed event. This results in a density increase at the loop apex, faciliating the onset of thermal instability in the loop. A few outstanding questions still remain though. By what fraction does the density of the unstable loop increase after the onset of the heating? Where do the flows responsible for the redistribution of the mass in the loop originate; in the vicinity of the reconnection site or at the footpoints in the chromosphere?  

The main difference between the rain event analysed in this work and the flare-driven coronal rain \citep{scullion_observing_2016} is the localisation of the impulsive heating leading to the thermal instability onset. In the flare-driven scenario, the impulsive heating is localised at footpoints of post-flare loops and is caused by non-thermal electrons accelerated at the reconnection site hitting the chromosphere. This leads to rapid chromospheric evaporation and the dense plasma quickly fills the loop triggering the thermal instability. In the event studied here however, the impulsive heating is localised at the reconnection site and dominates over any footpoint heating, as evidenced by soft X-Ray emission observed by RHESSI localised above the limb. There were no observable X-ray footpoint sources. The kink instability induced reconnection event itself also differs from the standard flare scenario in which a large number of post-flare loops form continuously during a lift-off of the reconnection site.

Numerical simulations of electron beam heating associated with solar flares currently cannot reproduce the coronal rain formation, suggesting that the lifetime of the electron beams is too short to sustain the prolonged chromospheric evaporation into the loop necessary for the loop to become thermally unstable \citep{reep_electron_2018}. It is therefore worth considering the possibility that short duration impulsive events can also act as a perturbation that triggers the onset of catastrophic cooling in a loop that is marginally thermally unstable.

The entry of the coronal rain condensations into the transition region and chromosphere is accompanied by small-scale brightenings which can be observed in transition region and in the coronal channels. The emission in different transition region and chromospheric channels is cospatial to a large extent, with most of the condensations observable in the IRIS Si IV and Mg II K SJI data, and in the SDO/AIA 304 {\AA} channel. This implies that the structure of the condensation plasma is to a certain degree multi-thermal, similar to the coronal rain observed in quiescent coronal loops \citep{antolin_multithermal_2015, kohutova_analysis_2016}. The speeds of the individual coronal rain blobs measured to be $\sim$ 60 km s$^{-1}$ are comparable to those observed in the quiescent case \citep{antolin_transverse_2011, kohutova_analysis_2016}. This is smaller than the maximum free-fall speed of 150 km s$^{-1}$ estimated for a loop of this size. Hence the coronal rain condensations in the studied event move with less than free-fall speeds, which is a commonly observed phenomenon \citep[e.g.][]{antolin_transverse_2011, antolin_observing_2012, kohutova_analysis_2016, schad_neutral_2018}.

In this work we have shown that the magnetic reconnection can be directly responsible for the thermal instability onset and subsequent condensation formation through the impulsive localised heating associated with this process, and not just for downfalling motion of condensations due to reconnection induced topology changes, as previously shown by \citet{li_coronal_2018}. We have further shown that the thermal instability-inducing heating can therefore act at any location along the loop and not just at the footpoints, as previously assumed. Furthermore, coronal rain formation can be triggered by an impulsive one-off event, in addition to being caused by a sustained prolonged heating assumed in the previous models. It is unclear, however, whether the reconnection-induced changes in magnetic topology, such as the change in the loop length play a role in the instability onset and condensation formation in the studied event. It has been shown by numerical simulations of prominence formation that this is indeed possible; initially thermally stable short loops can reconnect and become longer, which leads to a decrease in the ratio of heating scale-height to loop length and to the loops becoming thermally unstable as a result \citep{kaneko_reconnection_2017}.

Ideally a more comprehensive observational study is needed to see how common the reconnection-induced coronal rain is in the corona. A significant drawback linked to performing a large-scale statistical study is a lack of a single direct evidence of the reconnection taking place and hence having to rely on multiple indirect reconnection signatures. These include the evolution of the large-scale magnetic field topology inferred from PFSS extrapolation; imaging data taken in coronal, transition region, and chromospheric wavelengths; the presence of fast reconnection outflows; and evolution of emission in different wavelengths, which can be used to infer the thermal evolution of the plasma and therefore can be used as evidence of reconnection-induced Ohmic heating. Not all of these are present at the same time, which means the observations of individual events have to be treated on a case-by-case basis. 

In order to understand the constraints on the magnetic reconnection location that is capable of triggering the thermal instability, multi-dimensional MHD simulations are necessary. They are also needed to obtain the estimate of the total heat flux generated during the reconnection process and how this is linked to the length of the instability timescale and to the overall mass of condensation plasma formed as a result. The estimates of the total mass fraction of the coronal loop plasma that becomes thermally unstable obtained from the observations can then be used to constrain the total heating generated during a real reconnection event.

\begin{acknowledgements}
This research was supported by the Research Council of Norway through its Centres of Excellence scheme, project no. 262622. E.V. acknowledges financial support from the UK STFC on the Warwick STFC Consolidated Grant ST/L000733/I. IRIS is a NASA small explorer mission developed and operated by LMSAL with mission operations executed at NASA Ames Research Center and major contributions to downlink communications funded by the Norwegian Space Centre (NSC, Norway) through an ESA PRODEX contract. The SDO/AIA data are available courtesy of NASA/SDO and the AIA, EVE, and HMI science teams.
\end{acknowledgements}

\bibliography{CR_reconnection}

\end{document}